\documentclass[twocolumn,prl,showpacs,floatfix,superscriptaddress]{revtex4}

\usepackage{amssymb}
\usepackage{graphicx}

\begin{document}

\title{The Quantum Hall Transition in Real Space: From Localized to Extended
States}
\author{K. Hashimoto}
\email{hashi@mail.tains.tohoku.ac.jp}
\affiliation{Institute of Applied Physics, Hamburg University, Jungiusstra$\beta $e 11,
D-20355 Hamburg, Germany}
\affiliation{Department of Physics, Tohoku University, Sendai 980-8578, Japan}
\affiliation{ERATO Nuclear Spin Electronics Project, Sendai 980-8578, Japan}
\author{C. Sohrmann}
\affiliation{Department of Physics and Center for Scientific Computing, University of
Warwick, Gibbet Hill Road, Coventry CV4 7AL, UK}
\author{J. Wiebe}
\affiliation{Institute of Applied Physics, Hamburg University, Jungiusstra$\beta $e 11,
D-20355 Hamburg, Germany}
\author{T. Inaoka}
\affiliation{Department of Physics and Earth Sciences, University of the Ryukyus, 1
Senbaru, Nishihara, Okinawa 903-0213, Japan}
\author{F. Meier}
\altaffiliation{Present address: Department of Physics, Cornell University, Ithaca NY 14853, USA}
\affiliation{Institute of Applied Physics, Hamburg University, Jungiusstra$\beta $e 11,
D-20355 Hamburg, Germany}
\author{Y. Hirayama}
\affiliation{Department of Physics, Tohoku University, Sendai 980-8578, Japan}
\affiliation{ERATO Nuclear Spin Electronics Project, Sendai 980-8578, Japan}
\author{R. A. R\"{o}mer}
\affiliation{Department of Physics and Center for Scientific Computing, University of
Warwick, Gibbet Hill Road, Coventry CV4 7AL, UK}
\author{R. Wiesendanger}
\affiliation{Institute of Applied Physics, Hamburg University, Jungiusstra$\beta $e 11,
D-20355 Hamburg, Germany}
\author{M. Morgenstern}
\affiliation{II. Institute of Physics B, RWTH Aachen University, Aachen,
D-52056, Germany}
\affiliation{JARA-Fundamentals of Future Information Technology}
\date{\today }

\begin{abstract}
Using scanning tunneling spectroscopy in ultra-high vacuum at low
temperature ($T$ = 0.3 K) and high magnetic fields ($B \leq $
12 T), we directly probe electronic wave functions across an integer quantum Hall
transition. In accordance with
theoretical predictions, we observe the evolution from localized drift
states in the insulating phases to branched extended drift states at the
quantum critical point. The observed microscopic behavior close to the extended state
indicates points of localized quantum tunneling, which are considered to be decisive for a
quantitative description of the transition.
\end{abstract}

\maketitle

Two-dimensional electron systems (2DES) are pa\-ra\-digms for quantum phase
transitions \cite{Sachdev}. For example, they exhibit
metal-insulator, integer quantum Hall (IQH), and
superconductor-insulator transitions \cite{Shangina}, all described
fairly successfully by a universal percolation-type model \cite{Dubi}. Some
macroscopic aspects of this percolation have been verified, e.g. by
transport experiments revealing critical exponents for the divergence of the
localization length \cite{Koch}. However, the most fundamental aspect, i.e.\
how wave functions change at the transition, has not been probed
experimentally. Here, we directly observe the wave functions at the IQH transition taking
advantage of its tunability by a magnetic ($B$) field.

The microscopic description of IQH transitions is theoretically well established
\cite{Joynt, Ando, Kramer}. The 2DES in $B$-fields 
exhibits discrete kinetic energies which are called Landau levels (LL). The corresponding states are
subject to random potential disorder, which has primarily a semiclassical effect \cite{Joynt}: the electrons perform a fast
cyclotron rotation within the electrostatic disorder which leads
to additional drift motion along the equipotential lines \cite{Mirlin}.
Quantum mechanically, so-called drift states meander along equipotential
lines with a width of about the cyclotron radius
$r_{c}$ \cite{Ando}. If the potential energy of the state is low (high), the drift states are closed trajectories around potential minima (maxima), i.e.\ they are localized and represent insulating electron
phases. In the center of a LL, the adjacent trajectories
merge at the saddle points of the potential leading to an extended state
traversing the whole sample. It is known that this state is the quantum
critical state of the IQH transition and responsible for the finite
longitudinal resistance between quantized values of the Hall conductance
\cite{Joynt, Ando, Kramer, Evers}.

Several experiments have addressed microscopic aspects of QH
transitions \cite{MMbook,Ilani}.
However, none of these techniques
revealed sufficient lateral resolution to image the drift states.
Only scanning tunneling spectroscopy (STS) using the adsorbate-induced 2DES at n-InAs(110) surfaces
found localized drift states within a LL tail \cite{MM2003}.
This paves the path to observe the decisive transition from localized to
extended states directly. Localized states in $B$-field have also been observed
on graphite, which, however, does not exhibit a pure 2DES \cite{Niimi}.

Here, we probe the local density of states (LDOS) of the
adsorbate-induced 2DES on n-InSb(110) by STS.
The 2DES is prepared by depositing
$0.01$ monolayer of Cs atoms on cleaved n-InSb(110) at $T=30$ K and $p \sim 10^{-10}$ mbar \cite{EPAPS,Betti}.
STS is performed in-situ at $T$ = 0.3 K and $B \le 12$ T \cite{Wiebe} using a W-tip carefully selected 
by trial and error 
to exhibit negligible tip-induced band-bending \cite{EPAPS,Dombrowski}. 
The tip is stabilized at current $I_{\rm stab}$ and voltage $V_{\rm stab}$, prior to measuring d$I/$d$V$ 
as a function of sample voltage $V_s$ directly by lock-in technique with modulation voltage $V_{\rm mod}$.

\begin{figure}[tb]
\includegraphics{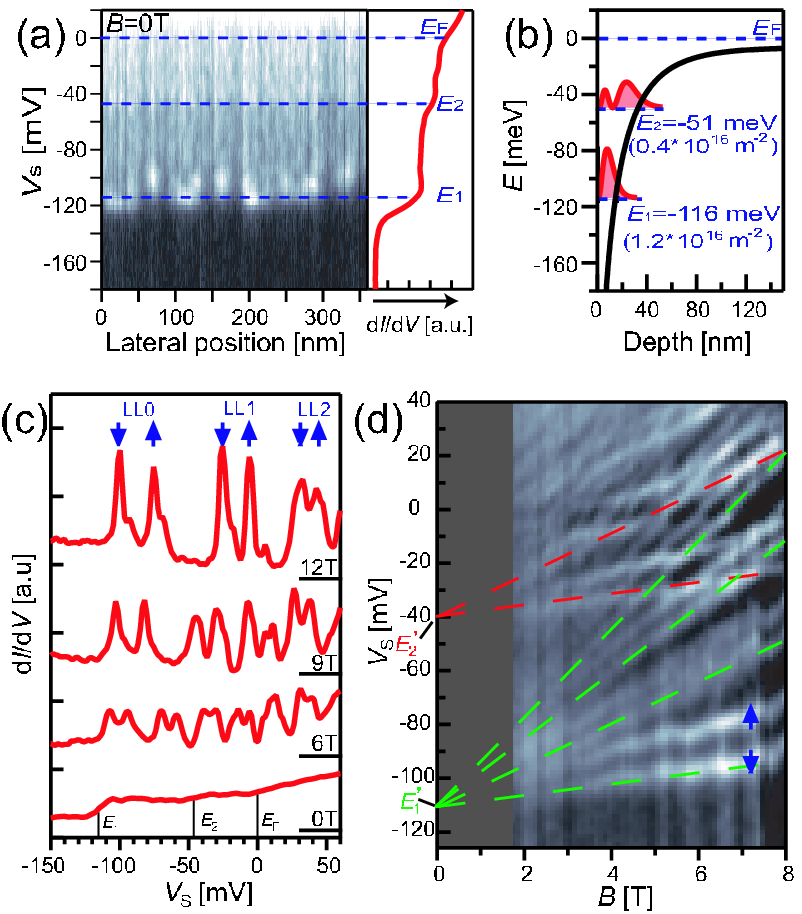}
\caption{(a) Color scale plot of d$I/$d$V(V_s)$ measured along a straight line at $B=0$ T (left); corresponding
spatially averaged d$I/$d$V(V_s)$ (right); dashed lines  mark the Fermi level
$E_{F}$ and $i$th subband energies $E_{i}$ ($i = 1,2$); $V_{\rm stab}=150$ mV, $I_{\rm stab}=0.13$ nA,
$V_{\rm mod}=2$ mV. (b) Self-consistently calculated band bending at the InSb surface (black line) and 
electron-density distribution of $i$th subband (red areas); $E_i$ and
electron areal densities $n_i$ are marked. (c) d$I$/d$V(V_s)$
at different $B$ recorded at the same lateral position;
$E_{i}$, $E_{F}$, ($dI/dV$ = 0)-lines on the right, LLs and spin directions are marked;
$V_{\rm stab}=150$ mV, $I_{\rm stab}= 0.13$ nA, $V_{\rm mod} = 2.0$ mV (0 T),
1.3 mV (6 T), 1.0 mV (9 T), 0.9 mV (12 T) (d) Experimentally determined Landau fan
diagram; green (red) dashed lines: spin-down LLs of the 1st (2nd) subband; arrows:
spin-up levels of the lowest LL; $V_{\rm stab}=150$ mV, $I_{\rm stab}=0.10$ nA, $V_{\rm mod}=1.5$ mV.}
\label{fig-1}
\end{figure}

First we sketch the general properties of the 2DES. 
Figure \ref{fig-1}(a) shows a color scale plot of the {\em local} differential conductivity
d$I/$d$V(V_{s})$ of the 2DES measured along a
straight line at $B = 0$ T and
the corresponding spatially averaged d$I/$d$V$
curve. They, respectively, represent the energy
dependence of the LDOS (energy-position plot) and of the macroscopically
averaged LDOS, i.e.\ the DOS \cite{MM2002}. The LDOS of the 2DES shows
two apparent boundaries coinciding with two step-like features in the
DOS at $V_{s}$ = -115 and -47 meV, which represent the first ($E_{1}$) and second (%
$E_{2}$) subband edges as indicated in Fig. \ref{fig-1} (a).
The subband energies are excellently reproduced by
a self consistent calculation [Fig.\ \ref{fig-1}(b)]
\cite{EPAPS,Abe}. The irregularity of the onset at \textit{E%
}$_{1}$ 
is a signature of the potential disorder \cite{MM2002}.
Figure 1(c) shows a set of d$I$/d$V$ curves measured at the
same position at different $B$. Starting at $B = 6$ T,
the d$I$/d$V$ curves exhibit distinct LLs with a pronounced
twofold spin splitting. The LLs are separated by regions 
of d\textit{I}/d\textit{V} $\approx $
0 evidencing complete quantization.
Indeed, spin-resolved IQH plateaus up to filling factor six 
were recently observed by
magnetotransport on an adsorbate-induced 2DES on InSb(110) \cite{Masutomi}. 
Repeating the measurement using smaller $B$-field steps highlights the continuous evolution of the spin-split LLs, i.e.\
the LL fan diagram in Fig.\ \ref{fig-1}(d). The green (red) dashed lines
mark the four (two) spin-down LLs of the 1st (2nd) subband. The accompanying
spin-up LLs are visible at higher energies as marked by blue
arrows for the lowest LL. Partly, LLs of different subbands cross without
anticrossing indicating orthogonality and, thus, negligible interaction
between the subbands.
%
From the peak distances, we deduce 
the effective mass 
$m^{\ast }$ and 
the absolute value of the $g$-factor
$|g|$ via the separation of LLs 
$\hbar |e|B/
m^{\ast }$ ($\hbar $: Dirac's constant, $e$: electron charge) and of spin levels
$|g|\mu _{B} B$ ($\mu _{B}$: Bohr magneton).
For the lowest peaks at $B=6$ T,
we find $m^{\ast }/m_{e}$ = 0.019$\pm $0.001 ($m_{e}$:
free-electron mass)  and $|g|=39\pm 2$. This is close to the
known values at the band edge $m^{\ast }/m_{e}$ = 0.014
and $|g| = 51$ with slight
deviations due to non-parabolicity and energy dependent spin-orbit coupling as known for InSb \cite{Merkt,Chantis}.

\begin{figure*}[tb]
\includegraphics{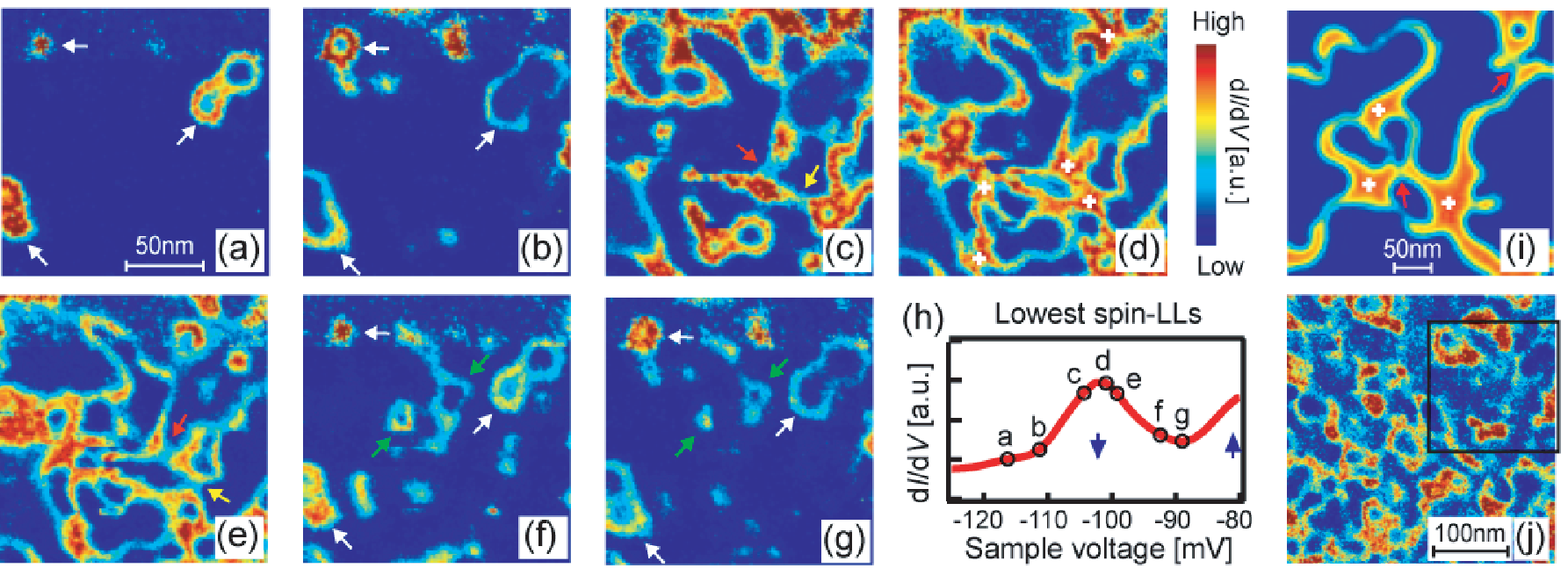}
\caption{LDOS of lowest LL. (a)-(g) Measured d$I$/d$V(x,y)$;
$B = 12$ T, $V_s = -116.3$ mV (a), -111.2 mV (b), -104.4 mV (c), -100.9 mV
(d), -99.2 mV (e), -92.4 mV (f), -89.0 mV (g), $I_{\rm stab} = 0.1$ nA, $V_{\rm stab} =
150$ mV, $V_{\rm mod} = 1.0$ mV; same d$I/$d$V$-color scale in each image;
white (green) arrows in (a), (b), (f), (g) mark drift states encircling potential minima (maxima);
red, yellow arrows in (c), (e) mark tunneling connections
existing at identical positions; crosses in (d) mark extended
LDOS areas at saddle points. (h) Spatially averaged d$I$/d$V$ curve with
circles at the $V_s$ used in (a)-(g). (i) Calculated LDOS at the
center of LL0$\downarrow$ at $B = 12$ T; red arrows mark tunneling
connections at the saddle points; white crosses mark extended areas. (j) d$I$/d$V$
image close to the center of LL0$\uparrow$ at $B = 6$ T, $V_s = -99$ mV;
image includes the area of (a)-(g) within the marked rectangle.}
\label{fig:wide}
\label{fig-2}
\end{figure*}

Figures \ref{fig-2}(a)-(g) show our central result, the LDOS across an IQH transition, i.e.\ d$I/$d$V$ images recorded in the lowest  spin-down LL of the 2DES
(LL0$\downarrow$) at $B= 12$ T.
The corresponding, spatially averaged d$I/$d$V$ curve is shown in Fig.\ \ref{fig-2}(h).
The continuous change of the LDOS with energy is available as
an EPAPS movie \cite{EPAPS}. The LDOS in the low-energy tail of LL0$\downarrow$ is shown in Fig.\ \ref{fig-2}(a).
It exhibits spatially isolated closed-loop
patterns with averaged full width at half maximum (FWHM) $\simeq$ 6.9 nm close to \textit{r}$_{c}$ = 7.4 nm. 
Thus, we attribute the
closed patterns to localized drift states of LL0$\downarrow$ aligning along
equipotential lines around a potential minimum. Accordingly, at slightly
higher energy [Fig.\ \ref{fig-2}(b)], the area encircled by the drift states increases
indicating that the drift states probe a longer equipotential line at higher
energy within the same valley. In contrast, the ring patterns at the
high-energy tail, marked by green arrows in Fig.\ \ref{fig-2}(f), (g), encircle an
area decreasing in size with increasing voltage. These states are attributed to
localized drift states around potential maxima. Notice that the structures
in Fig.\ \ref{fig-2}(a) and (b) appear nearly identically in Fig.\ \ref{fig-2}(f) and (g) as marked
by white arrows. The latter structures are the LL0$\uparrow$ states
localized around potential minima, which energetically overlap with the
high-energy LL0$\downarrow$ states localized around potential maxima. When the
voltage is close to the LL center [Fig.\ \ref{fig-2}(c), (e)], adjacent drift states
coalesce and a dense network is observed directly at the LL center [Fig.\ \ref{fig-2}(d)]. This
is exactly the expected behavior of an extended drift state at the IQH transition
\cite{Ando,Kramer}.  Fig.\ \ref{fig-2}(i) shows a calculated extended state at $B = 12$ T in a  2DES of InSb \cite{InSb} 
\cite{Sohrmann} and the potential disorder provided by the known dopant density of the sample \cite{EPAPS}. 
Good qualitative agreement with the measurement is achieved supporting the interpretation of the coalesced LDOS patterns as extended states.
Figure \ref{fig-2}(j) shows another extended state recorded on a larger area at different $B$ 
demonstrating that the coalesced patterns is not restricted to small length scale.

Interestingly, the drift states around the LL center
indicate quantum tunneling at the saddle points. Within the
classical percolation model, the adjacent drift states are connected at
a singular energy at each saddle eventually leading to a localization
exponent $\mathit{\nu }$ = 4/3. However, experimental \cite{Koch}, and
numerical \cite{Kramer} results for IQH transitions revealed $\mathit{\nu
\approx }$ 2.3. The discrepancy is attributed to quantum tunneling between
classically localized drift states \cite{Ando, Kramer}. Tunneling
connections are indeed
visible in our data spreading over an energy range of 5 meV. As an example, the red (yellow) arrows in Fig.\ \ref{fig-2}(c)
and (e) mark the same connection point at \textit{V}$_{s}$ = -104.4 mV and
-99.2 mV. LDOS is faintly visible at both positions in both images and
sharply rotates by about 90$^\circ$
 between the images. The reason is simply that the tunneling
interconnection mediates between valley states at low energies and between hill states at high energies, which
are connected via two nearly orthogonal lines. Such weak links are
reproduced by the calculation as marked by red arrows in Fig.\ \ref{fig-2}(i)
(see also Fig.\ S3 of \cite{EPAPS}). Note that the intrinsic energy resolution of
the experiment is 0.1 meV \cite{Wiebe}, while peaks in the LL fan diagram
exhibit a FWHM of 2.5 meV probably due to life-time effects. Thus, broadening due to
the energy resolution can hardly account for intensity
at the saddles within an energy range of 5 meV. Moreover, it cannot explain the change of
orientation. Another intriguing observation are the LDOS areas
larger than $r_c$ around the saddles. They are again visible in experiment [crosses in Fig.\ \ref{fig-2}(d)] and calculation
[crosses in Fig.\ \ref{fig-2}(i)] and are
probably due to the flat potential at the saddles leading to slow drift
speed and, thus, extended LDOS intensity. Notice that both, the spreading of LDOS intensity at the saddles 
in energy and position, is consistent with
previous quantum mechanical calculations \cite{Ando, Kramer}.

\begin{figure}[tb]
\includegraphics{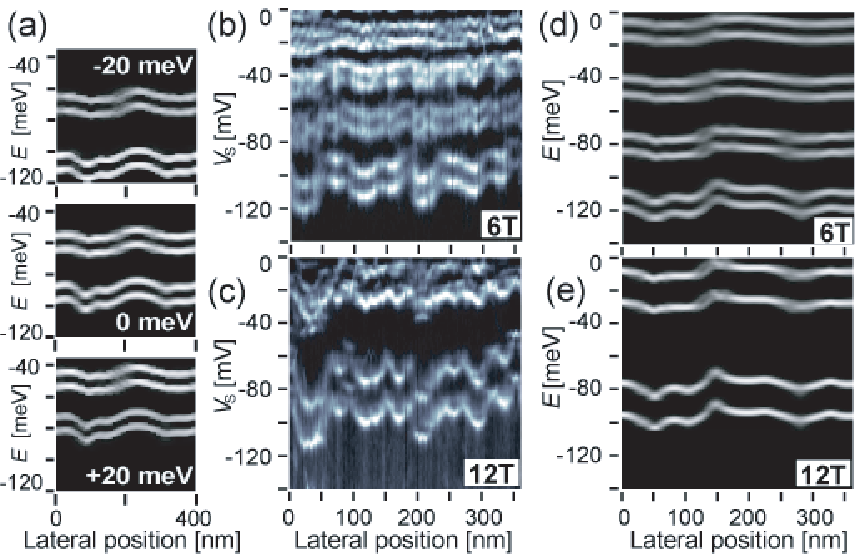}
\caption{(a) Calculated LDOS$(E,x)$ at $B= 6$ T; $V_{\rm tip} = 0$ and $\pm20$ meV as marked; $m^{\ast }/m_{e} :=0.02$, $|g| := 28$ \cite{InSb} (b), (c) Measured d$I$/d$V(V_s,x)$ at 
$B = 6$ and $12$ T; $x$: lateral position along straight line; $I_{\rm stab}$ = 0.13 nA,
$V_{\rm stab}$ = 150 mV, $V_{\rm mod}$ = 1.3 mV (b), 0.9 mV (c). 
(d), (e) Calculated LDOS$(E,x)$ at $B = 6$ and $12$ T; $V_{\rm tip}$ = 0 meV.} 
\label{fig-3}
\end{figure}

Finally, we discuss the possible influence of the tip. It is known that a mismatch of tip and
surface potential leads to band bending within the sample \cite{Dombrowski,MM2002}.
To avoid this, we used only W-tips exhibiting a
minimum of tip-induced band-bending. By analyzing the $dI/dV$ data at $B=0$ T with and without adsorbates, 
we can safely  rule out a work function mismatch between tip and sample larger than 15 meV \cite{Dombrowski,MM2002,EPAPS}.
The applied $V_{s}$ leads to an additional tip-induced band bending
with a lever arm of ten as determined from experiments described in \cite{EPAPS}.
Thus, we get an additional band bending of less than 12 meV.
The influence of such small band bendings is tested by the theoretical
calculations of a disordered 2DES in $B$-field \cite{EPAPS,InSb,Sohrmann}.
The tip-induced potential is added to the disorder potential of the 2DES as a Gaussian with 50 nm FWHM
and amplitude $|V_{\rm tip}|< 20$ meV \cite{Dombrowski,MM2002}.
The LDOS is calculated at each lateral tip position and its intensity below the tip is
displayed as a function of tip position. Fig.\ \ref{fig-3}(a) shows the resulting energy-position plots for three different $V_{\rm tip}$ 
at $B=6$ T. Obviously, $V_{\rm tip}$ leads only to a rigid energetic shift. 
This result is robust to changes in $B$ and FWHM.  Thus, the spatial
dependence of the LDOS patterns is not influenced by a small potential mismatch $|V_{\rm tip}| < 20$ meV.

To substantiate this result, Figs. \ref{fig-3}(b)-(e) show experimentally determined voltage-position plots 
in comparison with calculated energy-position plots \cite{EPAPS,InSb} at $B$ = 6 and 12 T. Spin-split LLs fluctuating in energy are visible as
meandering pairs of lines.
%
%
The theoretical calculations do not require free parameters by neglecting the tip-induced potential.
They reproduce the experimental trends of spatially fluctuating spin resolved LLs. 
The fluctuation length is
80 \% larger and the fluctuation amplitude 40 \% lower than in the
experiment. This is most likely due to the neglected inhomogeneous charge
distribution of the adsorbates, which would increase the potential disorder given by the bulk dopants.
Indeed, as shown in Fig.\ S1 of \cite{EPAPS}, the spatial distribution of adsorbates is 
related to the resulting drift states, and the correspondence is improved by adding surface charges to the
calculation (not shown).
However, since many details of the adsorbate potential are not
known, we omit its inclusion, thereby avoiding the use
of further parameters. Note that the parameter-free calculation and the
experiment quantitatively exhibit the same trends with $B$-field and
energy, i.e.\ a 30 \% increase of fluctuation amplitude between 6 and 12 T
and a 60 \% decrease of fluctuation amplitude between LL0 and LL3 at $
B = 6$ T. Both effects are explained straightforwardly by the fact that
drift states can only probe potential fluctuations down to length scales of
$r_{c} = \sqrt{\left(2n+1\right) \cdot \hbar / \left(|e|B\right) }$ ($n$: LL index). 

In summary, our STS experiments provide the first direct observation of wave functions across an IQH transition.
They reveal the development from localized to extended states including indications of the theoretically predicted
quantum tunneling at saddle points of the disorder potential.
We probe states well away from the Fermi level, and thus artificially exclude the influence of electron-electron
interaction, thereby considering a theoretically pure IQH transition, which is regarded as a hallmark in the theoretical description \cite{Ando,Kramer,Mirlin,Evers}. Indeed, single-particle calculations can largely reproduce our experimental
results. In principle, the experiments can be extended to 2DES states at $E_F$ by using p-type samples. Thus, we could add the electron-electron interaction, which reveals a wealth of further quantum phase transitions, in future experiments \cite{Evers,Ilani}.
However, already the present results go far beyond previous results \cite{MM2003,Niimi} by providing the states at the quantum phase transition, a LL fan diagram and a detailed theoretical reproduction of the observed experimental LDOS features.

We thank F. Evers, H. Akera, G. Bauer, M. B. Santos, D. Haude, F. Marczinowski, S. v. Oehsen, and G. Meier for helpful discussions
and the DFG program \textquotedblleft Quantum-Hall
systems\textquotedblright\ as well as SFB
508-B4 for financial support. 
Part of the calculations were performed at ISIC, Iwate University and Cyberscience Center, Tohoku
University as well as at the UK National Grid Service.


\begin{thebibliography}{99}
\bibitem{InSb} We use constant values $m^{\ast }/m_{e} =0.02$, $|g| = 28$ determined by energetically averaging the Landau and spin level separation
in Fig. 3(b).
\end{thebibliography}
\end{document}